# Insulating Parent Phase and Distinct Doping Evolution to Superconductivity in Single-Layer FeSe/SrTiO$_3$ Films

Yong Hu[1,2#], Yu Xu[1,2#], Yi-Min Zhang[3#], Qing-Yan Wang[1,2], Shao-Long He[1], De-Fa Liu[1], Ai-Ji Liang[1], Jian-Wei Huang[1,2], Cong Li[1,2], Yong-Qing Cai[1,2], Ding-Song Wu[1,2], Guo-Dong Liu[1,2,5], Fang-Sen Li[3], Jia-Qi Fan[3], Guan-Yu Zhou[3], Lili Wang[3], Can-Li Song[3*], Xu-Cun Ma[3], Qi-Kun Xue[3,6], Zu-Yan Xu[4], Lin Zhao[1,2*] and X. J. Zhou[1,2,5,6*]

[1]National Lab for Superconductivity, Beijing National Laboratory for Condensed Matter Physics, Institute of Physics, Chinese Academy of Sciences, Beijing 100190, China

[2]University of Chinese Academy of Sciences, Beijing 100049, China

[3]State Key Lab of Low-Dimensional Quantum Physics, Department of Physics, Tsinghua University, Beijing 100084, China

[4]Technical Institute of Physics and Chemistry, Chinese Academy of Sciences, Beijing 100190, China

[5]Songshan Lake Materials Laboratory, Dongguan 523808, China

[6]Beijing Academy of Quantum Information Sciences, Beijing 100193, China

[#]These authors contribute equally to the present work.

[*]Corresponding author: lzhao@iphy.ac.cn, clsong07@mail.tsinghua.edu.cn and XJZhou@iphy.ac.cn

**The single-layer FeSe/SrTiO$_3$ (FeSe/STO) films have attracted much attention because of their simple crystal structure, distinct electronic structure and record high superconducting transition temperature ($T_c$). The origin of the dramatic $T_c$ enhancement in single-layer FeSe/STO films and the dichotomy of superconductivity between single-layer and multiple-layer FeSe/STO films are still under debate. Here we report a comprehensive high resolution angle-resolved photoemission spectroscopy and scanning tunneling microscopy/spectroscopy measurements on the electronic structure evolution with doping in single-layer and multiple-layer FeSe/STO films. We find that the single-layer FeSe/STO films have a distinct parent phase and a route of doping evolution to superconductivity that are fundamentally different from multiple-layer FeSe/STO films. The parent phase of the single-layer FeSe/STO films is insulating, and its doping evolution is very similar to that of doping a Mott insulator in cuprate superconductors. In multiple-layer FeSe/STO films, high-$T_c$ superconductivity occurs by suppressing the nematic order in the parent compound with electron doping. The single-layer FeSe/STO films represent the first clear case in the iron-based superconductors that the parent compound is an insulator. Our observations of the unique parent state and doping evolution in the single-layer FeSe/STO films provide key insight in understanding its record high-$T_c$ superconductivity. They also provide a new route of realizing superconductivity in iron-based superconductors that is common in high temperature cuprate superconductors.**

The FeSe-based superconductors have attracted much interest recently because they provide an ideal system to study the interplay between magnetism, nematicity and superconductivity. The bulk FeSe has a simple crystal structure consisting solely of the basic building block of the FeSe-based superconductors [1]. It shows a structural transition around 90 $K$ without a magnetic transition, and its $T_c$ of ~8 $K$ at ambient pressure can be enhanced to 37~40 $K$ under high pressure [2, 3]. Doping electrons into bulk FeSe through gating or potassium deposition leads to a $T_c$ increase up to 48 $K$ [4–7]. Intercalation of bulk FeSe gives rise to $A_xFe_{2-y}Se_2$ (A=K, Rb, Cs, Tl and etc.) superconductors with a $T_c$ of 30-55 $K$ [8–14] and (Li,Fe)OHFeSe superconductor with a $T_c$ of 42 $K$ [15]. One remarkable discovery is the observation of high temperature superconductivity with a $T_c$ above 65 $K$ in the single-layer FeSe films grown on the $SrTiO_3$ substrates [16–25]. The superconducting single-layer FeSe/$SrTiO_3$ (FeSe/STO) films exhibit a distinct electronic structure that consists of only electron pockets around the Brillouin zone corner without Fermi pockets near the zone center [17–19, 26, 27]. There remain some prominent issues to be resolved related to the superconductivity of FeSe/STO films. The first is the origin of the dramatic $T_c$ enhancement in the single-layer FeSe/STO films and whether it can be attributed to the interface effect or electron coupling with phonons in the STO substrate [16, 26, 28–30]. The second issue concerns the dichotomy between the single-layer and multiple-layer FeSe/STO films on why they behave differently upon carrier doping [16, 19, 31–33]. The third issue points to the parent phase of the superconducting single-layer FeSe/STO films and its doping evolution to superconductivity [18, 27, 34]. Answers to these questions are significant to further enhancing superconductivity and understanding the superconductivity mechanism in iron-based superconductors.

In this work, we report a comprehensive high resolution angle-resolved photoemission spectroscopy (ARPES) and scanning tunneling microscopy/spectroscopy (STM/STS) measurements on the electronic structure evolution with doping in single-layer and multiple-layer FeSe/STO films. Three doping procedures are employed: vacuum annealing, *in situ* potassium deposition to dope electrons and *in situ* selenium deposition to dope holes. We find that the parent phase of the single-layer FeSe/STO films is insulating. This is distinct from the nematic parent phase in multiple-layer FeSe/STO films and other iron-based superconductors. It represents the first clear case in the iron-based superconductors that the parent compound is an insulator. The doping evolution of single-layer FeSe/STO films is similar to doping a Mott insulator in high temperature cuprate superconductors. In multiple-layer FeSe/STO films, superconductivity occurs by suppressing the nematic order in the parent compound with electron doping. These observations point to a significant electron correlation in the single-layer FeSe/STO films and provide a new route of realizing superconductivity in iron-based superconductors which is similar to that in high temperature cuprate superconductors.

The single-layer and multiple-layer FeSe/STO films were prepared by the molecular beam epitaxy (MBE) method [16, 35], and were characterized by *in situ* STM that is connected directly to the MBE chamber (See Methods in Supplementary). High resolution ARPES measurements were performed on the FeSe/STO films with different doping levels (See Methods). To change the doping level of the samples, we used three different ways: vacuum annealing, potassium deposition and selenium deposition. The measurements were repeated several times and the results are highly reproducible.

Figure 1 shows Fermi surface and band structure evolution of the single-layer FeSe/STO film at different vacuum annealing sequences. A perfect single-layer FeSe/STO film was made in the MBE chamber, as characterized by STM (Fig. 1g1), and covered by amorphous Se deposition. Then the sample was transferred into the annealing chamber that is directly connected to the ARPES measurement chamber. The sample was annealed in ultra-high vacuum at different temperatures and for different times (see Fig. S1a in Supplementary); its electronic structure was measured after each annealing sequence. Fig. 1a shows the Fermi surface evolution with annealing at different sequences (A1-A7). The corresponding band structure evolution along three momentum cuts crossing Γ, $M_2$ and $M_3$ are shown in Fig. 1b, 1c and 1d, respectively. Two distinct Fermi surface topologies are observed during the annealing process (Fig. 1a). For the initial sequence A1, the Fermi surface displays four strong spots around M points. For the last sequence A7, it shows circular electron pockets around M points. The four-strong-spots Fermi surface gradually diminishes and the circular Fermi surface becomes dominant with annealing. They coexist in the intermediate annealing process (A2-A4). There are also two distinct sets of band structure observed in the annealing process. For the initial sequence A1, there is a hole-like band (marked as α in Fig. 1b) around Γ with its top about 10 *meV* below the Fermi level. Around M points, there are two hole-like bands (marked as $β_1$ and $β_2$ in Fig. 1d) and one flat band (marked as η in Fig. 1d) very close to the Fermi level. For the last sequence A7, there is a strong hole-like band (marked as A in Fig. 1b) around Γ with its top about 80 *meV* below the Fermi level, and an electron-like band (marked as D in Fig. 1b). Around M points, an electron-like band and a hole-like band with its top about 0.12 *eV* below the Fermi level (marked as C and B in Fig. 1d, respectively) are observed. There are also some additional bands with low spectral weight (A' in Fig. 1b and C' and B' in Fig. 1d) that may be attributed to the replica bands [26]. The two sets of bands are well separated and schematically shown in Fig. 1e and 1f for the bands near Γ and M points, respectively. With annealing, the first set of bands gets weaker and eventually disappears while the second set of bands emerges, gets stronger and finally becomes dominant. In the intermediate annealing process, these two sets of bands coexist. The electronic structure evolution during the vacuum annealing process is similar to that observed before where these two sets of distinct band structure were assigned to two separate phases, named as N and S phases, of the single-layer FeSe/STO films [18].

It has long been a puzzle on the nature of the N phase and S phase, their relationship and real space distribution in the single-layer FeSe/STO films [18, 22]. Although ARPES measurements suggest coexistence of the N and S phases in the vacuum annealing process of the single-layer FeSe/STO films [18], no signature of two coexisting phases are observed in STM measurements [36]. This urged us to reexamine on this issue. Recently, we have found that there can be a surprisingly large amount of excess Fe (~20%) in the single-layer FeSe/STO films [35]. We also found that, when depositing Se on the surface of single-layer FeSe/STO films, Se may react with the excess Fe to form the second FeSe-layer at an unexpectedly low temperature (150°C) which is much lower than the usual growth temperature of FeSe/STO films (~490 °C) [35]. When a perfect single-layer FeSe/STO film (Fig. 1g1) is capped with Se at room temperature and annealed at 150°C, instead of Se evaporation as usually expected, a large amount of islands (yellow regions in Fig. 1g2) are observed that represent the formation of the second FeSe-layer from reaction of Se with the excess Fe [35]. With further annealing at higher temperature and for longer time, the amount of the second FeSe-layer gradually decreases (Fig. 1g3-g6); they eventually disappear completely and the sample recovers to a perfect single-layer FeSe/STO film (Fig. 1g7) [35].

The samples we measured here and before [18] were capped with amorphous Se for sample protection, a popular method that is widely used in surface science [37, 38], when they are transferred from MBE preparation chamber to the ARPES measurement chamber. Since amorphous Se can be easily removed by annealing at a relatively low temperature, usually below 200°C [38], it was believed that the measured electronic structure of the single-layer FeSe/STO film after a mild annealing represents its intrinsic properties [18]. The unexpected discovery of the second FeSe-layer formation after Se deposition on single-layer FeSe/STO films and annealing at a relatively low temperature [35] provides a new understanding on the nature of the N and S phases in single-layer FeSe/STO films [18]. The Fermi surface and the band structure of the N phase are very similar to those of the double-layer FeSe/STO films [31]. The signal decrease of the N phase (Fig. 1a-d) with vacuum annealing and its eventual disappearance are in a good agreement with the decrease and final vanishing of the second FeSe-layer (Fig. 1g2-g7). These observations indicate that the signal of the N phase comes from the double-layer regions of the FeSe/STO films because of the second FeSe-layer formation due to Se deposition and annealing; it does not represent the intrinsic electronic structure of the single-layer FeSe/STO films [18]. In fact, it represents the nematic phase of the double-layer FeSe/STO films because the Fermi surface, band structure and their evolution with temperature resemble those of multiple-layer FeSe/STO films, bulk FeSe and other parent compounds of iron-based superconductors [19, 39, 40]. The signal of the S phase originates from the single FeSe-layer regions that represents intrinsic properties of the single-layer FeSe/STO films. Such an assignment resolves the puzzle why the coexistence of the N and S phases are not observed in single-layer FeSe/STO films by STM [36] although it was suggested from ARPES measurements [18].

The clear separation of signals from single-layer and double-layer FeSe/STO regions provides a good opportunity to study their electronic structure evolution with doping under the same vacuum annealing condition. With vacuum annealing, the electronic structure of the double-layer regions shows little change except that its signal gets weaker and eventually disappears. On the other hand, the signal of the single-layer regions emerges, gets stronger with vacuum annealing and finally becomes dominant. In the mean time, the electron pockets around M get larger, corresponding to an increase of electron doping (Fig. S3 in Supplementary). It is clear that, under the same vacuum annealing condition, the single-layer can be highly electron doped and becomes superconducting (Fig. S3 in Supplementary), but the electron doping of the double-layer regions remains quite small. These observations are consistent with the STM results where the single-layer and double-layer regions on the same sample surface exhibit disparate behaviors [16]. Our results indicate that the dichotomy originates from the difference of electron doping in single-layer and double-layer regions under the same annealing condition [31]. It is interesting to note that, for the A1 sequence sample (left-most panels in Fig. 1a-d), although the sample surface has nearly 80% of single-layer FeSe region, no trace of the signal from the single-layer is observed. When the signal of the single-layer FeSe regions becomes visible (second left-most panels in Fig. 1a-d for the A2 sequence sample), the doping level of the single-layer already reaches ~0.086 electrons per Fe (e/Fe). It is natural to ask what electronic structure of the single-layer FeSe is at lower electron doping.

By tuning the preparation conditions, we have successfully grown single-layer FeSe/STO films that is nearly free from excess Fe (see Fig. S4 in Supplementary). This makes it possible to investigate the electronic structure evolution with doping for the pure single-layer FeSe/STO films without the signal interference from the second layer, especially at the low doping levels. The sample was annealed in ultra-high vacuum at different temperatures and for different times (see Fig. S1c in Supplementary). Fig. 2 shows electronic structure evolution with vacuum annealing at different sequences (A1-A7). For the first three sequences A1-A3, there is no spectral weight detected at the Fermi level (Fig. 2a). The Fermi sur- face starts to emerge for sequence A4 with four electron pockets appearing near M points, corresponding to an electron doping of ~0.086 e/Fe. The electron pocket increases in size with further annealing (A5-A7 sequences in Fig. 2a), corresponding to an electron doping increase (Fig. 2e). For the sequences A1-A3, from the measured band structure near Γ and M points (left three panels in Fig. 2b-d), there are no spectral weight observed at the Fermi level but a spectral weight falling-off from high binding energy to the Fermi level; the spectral weight gradually shifts towards the Fermi level with increasing doping. This can be more clearly seen from the photoemission spectra (energy distribution curves, EDCs) in Fig. 2f. These results indicate that the single-layer FeSe/STO film is insulating in this low doping regime. For the sequences A4-A6, there appear clear hole-like bands near Γ point and electron-like bands near M points (Fig. 2b-d). But there remains an energy gap at the Fermi level which decreases in size from A4 to A6 samples (Fig. 2g). The observed gap is an insulating gap because it does not change its size with temperature [27]. This doping range (lower than ~0.092 e/Fe) is consistent with the previous result [27] that the samples are still insulating in this region. The sample A2 in Fig. 1 also lie in this region (Fig. S3 in Supplementary). For the sequence A7, when the doping level reaches ~0.106 e/Fe, it becomes superconducting (Fig. 2g). There is an insulator-superconductor crossover around the doping level of ~0.09 e/Fe in the single-layer FeSe/STO films [27]. The hole-doping into a superconducting perfect single-layer FeSe/STO film by successive Se depositions (see Fig. S8 in Supplementary) gives a fully consistent evolution picture as electron-doping into the parent phase by vacuum annealing presented here. The electronic structure evolution with doping (Fig. 2a-d) is also fully consistent with that from single-layer FeSe/STO regions in Fig. 1a-d. The insulating nature of the parent phase and its doping evolution are very similar to that in cuprate superconductors where insulator-superconductor transition is realized by doping a Mott insulator [27, 41].

The above results (Fig. 1 and Fig. 2) have established that the parent phase of the single-layer FeSe/STO films is insulating and superconductivity is realized by doping the insulating parent compound. The parent phase of double-layer FeSe/STO films is apparently not insulating; it is nematic because the Fermi surface (left-most panel in Fig. 1a), band structure (left-most panel in Fig. 1b-d) and their evolution with temperature resemble those of multiple-layer FeSe/STO films [19], bulk FeSe [40] and other parent compounds of iron-based superconductors [39]. The doping evolution of the double-layer FeSe/STO films is not accessible in Fig. 1 because its doping level changes little during the vacuum annealing process. To effectively tune the doping level in double-layer FeSe/STO films over a relatively large range, we turn to use potassium (K) deposition on the sample surface to introduce electrons [32, 33, 42]. Fig. 3 shows the evolution of Fermi surface and band structure with increasing potassium deposition for single-layer FeSe/STO film that has ~20% second FeSe- layer (similar to Fig. 1g2). Similar to Fig. 1, the electronic structure of the sample before potassium deposition (leftmost panels DK0 in Fig. 3a-d) originates predominantly from the double-layer FeSe regions. It is a nematic phase at a low temperature (~20 $K$) where the nematicity-induced splitting of the $d_{xz}$ and $d_{yz}$ bands near the M points is ~80 $meV$ and the nematic transition temperature is ~220 $K$ (Fig. S7 in Supplementary) [31, 39]. With the increase of potassium deposition, the Fermi surface evolves from a topology where four strong points are observed near M points which is characteristic of a typical nematic phase to the one with only electron pockets near M points (Fig. 3a). The estimated electron doping with potassium deposition is shown in Fig. 3e and the maximum doping level is ~0.072 e/Fe for the DK6 sample in Fig. 3a (also see

Fig. S6 in Supplementary). Over such an entire doping range, the low-energy signal near the Fermi level is dominated by the second FeSe-layer because the single-layer FeSe remains insulating in this doping range with weak signal (see Fig. 1, Fig. 2 and Fig. S8 in Supplementary) [27].

The doping evolution of the band structure for the double-layer FeSe/STO films follows a totally different route from that of the single-layer ones. With increasing potassium deposition, the hole-like band around Γ (marked as α in Fig. 3b DK0 panel) gradually shifts down from near the Fermi level to 80 *meV* below the Fermi level (DK1-DK6 in Fig. 3b and Fig. 3f). The electron-like band emerges near M point (Fig. 3c) and its bottom shifts down with increasing potassium deposition. This gives rise to an increase in the area of the electron-like Fermi surface near M points (Fig. 3a). With the increase of electron doping, the splitting between the $d_{yz}$ ($β_1$) and $d_{xz}$ ($β_2$) bands near M gets smaller and eventually disappears (Fig. 3d and 3g). Since this band splitting is associated with nematicity, it indicates that the nematicity is suppressed with increasing electron doping. At sufficiently high electron doping (~0.06 e/Fe), the nematicity is fully suppressed. In this case, the observed electronic structure of the double-layer FeSe (DK6 panels in Fig. 3a-d) becomes similar to that of the single-layer FeSe films (A7 panels in Fig. 1a-d, A4-A7 panels in Fig. 2a-d). We also prepared pure double-layer FeSe/STO films and investigated its doping evolution with potassium deposition (Fig. S9 in Supplementary); the results are consistent with that in Fig. 3.

We carried out STM/STS measurements to further examine on the nature of the parent phase and its doping evolution of the FeSe/STO films. The sample preparation and vacuum annealing procedures follow those we used in Fig. 1. Here we focus on the parent phase and low doping level since the doping evolution of single-layer FeSe/STO films at higher doping level has been studied before [36]. Fig. 4a shows the atomic resolution STM images of the sample surface after different annealing sequences, all showing 1×1 structure of the single- layer FeSe/STO films. The corresponding dI/dV curves for different annealing sequences are shown in Fig. 4b. With the advantage of measuring both occupied and unoccupied states, it is clear that the single-layer is insulating after the initial annealing sequence Q1; the full gap size is about 0.70 *eV*. The gap size gets smaller with annealing when electrons are gradually doped into the sample; it becomes ~80 *meV* for the sequence Q4. The gap shrinking and the spectral weight transfer towards the Fermi level are consistent with the ARPES results in Fig. 2. The much larger gap observed in STM/STS measurements (sequences Q1-Q3 in Fig. 4b) indicates that their doping levels are much smaller than those in Fig. 2 and closer to zero doping. Fig. 4c and Fig. 4d show detailed dI/dV curves for the single-layer and double-layer FeSe regions, respectively, on the same sample (inset of Fig. 4c) that corresponds to the sequence Q3 in Fig. 4a and Fig. 4b. While the single-layer FeSe is insulating without spectral weight at the Fermi level (Fig. 4c), the double-layer FeSe has considerable amount of spectral weight at the Fermi level (Fig. 4d). These results strongly support the above ARPES measurements that the single-layer and double-layer FeSe have fundamentally different nature of their parent phases.

In order to quantitatively keep track on the doping evolution of bands and compare their difference between single-layer and multiple-layer FeSe/STO films, we chose to plot the energy position of the band top for the hole-like bands near Γ (Fig. 5a) and band bottom for the electron-like bands near M (Fig. 5b) as a function of the electron concentration. These data come from different samples (single-layer, double-layer and 20-layer FeSe/STO films) with various doping methods (vacuum annealing, potassium deposition and Se deposition); they provide a consistent picture. For the single-layer FeSe/STO films, no signal can be detected from ARPES for the parent phase and at the low doping range. As soon as the Fermi surface and band structure become visible when the doping level is beyond a threshold (~0.07 e/Fe), the band top of the hole-like band near Γ and the band bottom of the electron-like band near M show little change with doping over a relatively large range (0.07~0.12 e/Fe). It is quite unusual because it does not follow a rigid band shift. Instead, the doping mainly changes the band shape of electron-like band near M that gives rise to an increase of the effective mass with increasing doping (Fig. 5c), consistent with the previous results [43]. For the double layer FeSe/STO films, the parent phase at zero doping is nematic. The band top of the hole-like band near Γ and the band bottom of the electron-like band near M exhibit an obvious shift to high binding energy with increasing electron doping. The band shift ceases at a critical doping level (~0.06 e/Fe) where the nematicity is fully suppressed. When the doping level becomes higher, the band position of those two bands shows little change with doping that is similar to the single-layer case. For the 20-layer FeSe/STO films (see Fig. S10 in Supplementary), its band evolution with doping is consistent with that of double-layer films. Fig. 5c summarizes the electron effective mass of all the FeSe/STO films deduced from electron-like band near M. It shows an overall increase with increasing electron doping, indicating an anomalous enhancement of correlation effects [43].

Our key findings of the present work can be summarized in the electronic phase diagrams for single-layer FeSe/STO films (Fig. 5d) and multiple-layer FeSe/STO films (Fig. 5e). The phase diagram of the single-layer FeSe/STO films is fundamentally different from that of double-layer and multiple-layer. As illustrated in Fig. 5d, the parent phase of the single-layer FeSe is insulating (with a full gap of ~0.7 *eV*, not shown in Fig. 5d). The gap shrinks with increasing doping and diminishes near ~0.09 e/Fe where there is an insulator-superconductor crossover [27]. Further increase of electron doping leads to an increase of the superconducting gap size and the corresponding $T_c$ [18]. On the other hand, for the double-layer and multiple-layer FeSe/STO films, the parent phase is nematic. The doping of electrons causes suppression of the nematic order and its disappearance at a critical value; it is ~0.06 e/Fe for the double-layer. When the doping level is beyond the critical doping, the Fermi surface only consists of electron pockets near M points, similar to the single-layer case and superconductivity can be realized in this doping range [32, 33, 42].

Our present results have established that the parent phase of the single-layer FeSe/STO films is insulating. This is the first clear case in iron-based superconductors that the parent phase is insulating because the parent compounds of other iron-based superconductors are known to be bad metals [44]. Its doping evolution is reminiscent of doping a Mott insulator in high temperature cuprate superconductors [45]. It provides a new route of realizing superconductivity in iron-based superconductors and establishes an intimate connection between iron-based superconductors and high temperature cuprate superconductors. It is important to understand the origin of the insulating nature in the parent phase of single-layer FeSe/STO films and whether the STO substrate plays a significant role in giving rising to the highly insulating state of the single-layer FeSe/STO parent phase. Such an understanding will help understand the dichotomy of parent phase and doping evolution between single-layer and multiple-layer FeSe/STO films. Our observations of the unique parent state and doping evolution in the single-layer FeSe/STO films provide key insight in understanding its record high-$T_c$ superconductivity. In addition to possible enhancement of $T_c$ through phonons in the STO substrate [26], our present work provides another way that $T_c$ enhancement may be realized through strong correlation effect induced by the STO substrate. The understanding of the underlying physics of the insulating nature in the parent phase, its unique doping evolution and the role they play in generating high temperature superconductivity in the single-layer FeSe/STO films will promote further investigations of superconductivity mechanism in iron-based superconductors.

**Acknowledgement**

We thank financial support from the National Key Research and Development Program of China (Grant No. 2016YFA0300300 and 2017YFA0302900), the National Natural Science Foundation of China (Grant No. 11888101), the Strategic Priority Research Program (B) of the Chinese Academy of Sciences (Grant No. XDB25000000), the National Basic Research Program of China (Grant No. 2015CB921000), the Youth Innovation Promotion Association of CAS (Grant No.2017013), and the Research Program of Beijing Academy of Quantum Information Sciences (Grant No. Y18G06).

**Author Contributions**

Y.H., Y. X. and Y.M.Z. contribute equally to this work. X.J.Z., L.Z., C.L.S. and Y.H. proposed and designed the research. Y.H., Y.X., Y.M.Z, F.S.L., J.Q.F., G.Y.Z., Q.Y.W., C.L.S., L.L.W. and X.C.M. contributed to MBE thin film preparation. Y.H., Y.X., L.Z., D.F.L., A.J.L., J.W.H., C.L., Y.Q.C., D.S.W., S.L.H., G.D.L., Z.Y.X. and X.J.Z. contributed to the development and

**Additional information**

Competing financial interests: The authors declare no competing financial interests.

**References**

[1] Hsu, F.-C. et al. Superconductivity in the PbO-type structure α-FeSe. *Proceedings of the National Academy of Sciences of the United States of America* **105**, 14262-14264 (2008).

[2] Medvedev, S. et al. Electronic and magnetic phase diagram of β-$Fe_{1.01}$Se with superconductivity at 36.7 *K* under pressure. *Nature Materials* **8**, 630-633 (2009).

[3] Sun, J.-P. et al. High-$T_c$ superconductivity in FeSe at high pressure: dominant hole carriers and enhanced spin fluctuations. *Physical Review Letters* **118**, 147004 (2017).

[4] Seo, J.-J. et al. Superconductivity below 20 *K* in heavily electron-doped surface layer of FeSe bulk crystal. *Nature Communications* **7**, 11116 (2016).

[5] Ye, Z.-R. et al. Simultaneous emergence of superconductivity, inter-pocket scattering and nematic fluctuation in potassium-coated FeSe superconductor. *arXiv*: 1512.02526 (2015).

[6] Song, C.-L. et al. Observation of double-dome superconductivity in potassium-doped FeSe thin films. *Physical Review Letters* **116**, 157001 (2016).

[7] Lei, B. et al. Evolution of high-temperature superconductivity from a low-$T_c$ phase tuned by carrier concentration in FeSe thin flakes. *Physical Review Letters* **116**, 077002 (2016).

[8] Guo, J.-G. Superconductivity in the iron selenide $K_x Fe_2 Se_2$ (0≤x≤1.0). *Physical Review B* **82**, 180520(R) (2010).

[9] Fang, M.-H. et al. Fe-based superconductivity with $T_c$=31 *K* bordering an antiferromagnetic insulator in (Tl,K)$Fe_x Se_2$. *Europhysics Letters* **94**, 27009 (2011).

[10] Ying, T.-P. et al. Observation of superconductivity at 30~46 *K* in $A_x Fe_2 Se_2$ (A = Li, Na, Ba, Sr, Ca, Yb, and Eu). *Scientific Reports* **2**, 426 (2012).

[11] Burrard-Lucas, M. et al. Enhancement of the superconducting transition temperature of FeSe by intercalation of a molecular spacer layer. *Nature Materials* **12**, 15-19 (2013).

[12] Hatakeda, T., Noji, T., Kawamata, T., Kato, M. & Koike, Y. New Li-ethylenediamine-intercalated superconductor $Li_x(C_2H_8N_2)_yFe_{2-z}Se_2$ with $T_c$=45 K. *Journal of the Physical Society of Japan* **82**, 123705 (2013).

[13] Shahi, P. et al. High-$T_c$ superconductivity up to 55 K under high pressure in a heavily electron doped $Li_{0.36}(NH_3)_yFe_2Se_2$ single crystal. *Physical Review B* **97**, 020508(R) (2018).

[14] Shi, M.-Z. et al. Organic-ion-intercalated FeSe-based superconductors. *Physical Review Materials* **2**, 074801 (2018).

[15] Lu, X.-F. et al. Coexistence of superconductivity and antiferromagnetism in $(Li_{0.8}Fe_{0.2})OHFeSe$. *Nature Materials* **14**, 325-329 (2014).

[16] Wang, Q.-Y. et al. Interface-induced high-temperature superconductivity in single unit-cell FeSe films on $SrTiO_3$. *Chinese Physics Letters* **29**, 037402 (2012).

[17] Liu, D.-F. et al. Electronic origin of high-temperature superconductivity in single-layer FeSe superconductor. *Nature Communications* **3**, 931 (2012).

[18] He, S.-L. et al. Phase diagram and electronic indication of high-temperature superconductivity at 65 K in single-layer FeSe films. *Nature Materials* **12**, 605-610 (2013).

[19] Tan, S.-Y. et al. Interface-induced superconductivity and strain-dependent spin density waves in $FeSe/SrTiO_3$ thin films. *Nature Materials* **12**, 634-640 (2013).

[20] Ge, J.-F. et al. Superconductivity above 100 K in single-layer FeSe films on doped $SrTiO_3$. *Nature Materials* **14**, 285-289 (2015).

[21] Zhang, Z.-C. et al. Onset of the Meissner effect at 65 K in FeSe thin film grown on Nb-doped $SrTiO_3$ substrate. *Science Bulletin* **60**, 1301-1304 (2015).

[22] Liu, X. et al. Electronic structure and superconductivity of FeSe-related superconductors. *Journal of Physics: Condensed Matter* **27**, 183201 (2015).

[23] Wang, L.-L., Ma, X.-C. & Xue, Q.-K. Interface high-temperature superconductivity. *Superconductor Science and Technology* **29**, 123001 (2016).

[24] Wang, Z., Liu, C., Liu, Y. & Wang, J. High-temperature superconductivity in one-unit-cell FeSe films. *Journal of Physics: Condensed Matter* **29**, 153001 (2017).

[25] Huang, D. & Hoffman, J. E. Monolayer FeSe on $SrTiO_3$. *Annual Review of Condensed Matter Physics* **8**, 311-336 (2017).

[26] Lee, J.-J. et al. Interfacial mode coupling as the origin of the enhancement of $T_c$ in FeSe films on $SrTiO_3$. *Nature* **515**, 245-248 (2014).

[27] He, J.-F. et al. Electronic evidence of an insulator-superconductor crossover in single-layer FeSe/SrTiO$_3$ films. *Proceedings of the National Academy of Sciences of the United States of America* **111**, 18501-18506 (2014).

[28] Xiang, Y.-Y., Wang, F., Wang, D., Wang, Q.-H. & Lee, D.-H. High-temperature superconductivity at the FeSe/SrTiO$_3$ interface. *Physical Review B* **86**, 134508 (2012).

[29] Zhang, S.-Y. et al. Role of SrTiO$_3$ phonon penetrating into thin FeSe films in the enhancement of superconductivity. *Physical Review B* **94**, 081116 (2016).

[30] Li, F. & Sawatzky, G. A. Electron phonon coupling versus photoelectron energy loss at the origin of replica bands in photoemission of FeSe on SrTiO$_3$. *Physical Review Letters* **120**, 237001 (2018).

[31] Liu, X. et al. Dichotomy of the electronic structure and superconductivity between single-layer and double-layer FeSe/SrTiO$_3$ films. *Nature Communications* **5**, 5047 (2014).

[32] Tang, C.-J. et al. Interface-enhanced electron-phonon coupling and high-temperature superconductivity in potassium-coated ultrathin FeSe films on SrTiO$_3$. *Physical Review B* **93**, 020507 (2016).

[33] Zhang, W.-H. et al. Effects of surface electron doping and substrate on the superconductivity of epitaxial FeSe films. *Nano Letters* **16**, 1969-1973 (2016).

[34] Kanayama, S. et al. Two-dimensional Dirac semimetal phase in undoped one-monolayer FeSe film. *Physical Review B* **96**, 220509 (2017).

[35] Hu, Y. et al. Identification of a large amount of excess Fe in superconducting single-layer FeSe/SrTiO$_3$ films. *Physical Review B* **97**, 224512 (2018).

[36] Zhang, W.-H. et al. Interface charge doping effects on superconductivity of single-unit-cell FeSe films on SrTiO$_3$ substrates. *Physical Review B* **89**, 060506 (2014).

[37] Wang, J. et al. Evidence for electron-electron interaction in topological insulator thin films. *Physical Review B* **83**, 245438 (2011).

[38] Virwani, K. et al. Controlled removal of amorphous Se capping layer from a topological insulator. *Applied Physics Letters* **105**, 241605 (2014).

[39] Liu, D.-F. et al. Common electronic features and electronic nematicity in parent compounds of iron-based superconductors and FeSe/SrTiO$_3$ films revealed by angle-resolved photoemission spectroscopy. *Chinese Physics Letters* **33**, 077402 (2016).

[40] Nakayama, K. et al. Reconstruction of band structure induced by electronic nematicity in an FeSe superconductor. *Physical Review Letters* **113**, 237001 (2014).

[41] Peng, Y.-Y. et al. Disappearance of nodal gap across the insulator-superconductor transition in a copper-oxide superconductor. *Nature Communications* **4**, 2459 (2013).

[42] Miyata, Y., Nakayama, K., Sugawara, K., Sato, T. & Takahashi, T. High-temperature superconductivity in potassium-coated multilayer FeSe thin films. *Nature Materials* **14**, 775-779 (2015).

[43] Wen, C.-H.-P. et al. Anomalouscorrelationeffectsanduniquephasediagramofelectron-doped FeSe revealed by photoemission spectroscopy. *Nature Communications* **7**, 10840 (2016).

[44] Stewart, G.-R. Superconductivity in iron compounds. *Reviews of Modern Physics* **83**, 1589-1652 (2011).

[45] Lee, P. A., Nagaosa, N. & Wen, X.-G. Doping a Mott insulator: Physics of high-temperature superconductivity. *Reviews of Modern Physics* **78**, 17-85 (2006).

[46] Norman, M.-R., Randeria, M., Ding, H. & Campuzano, J.-C. Phenomenology of the low- energy spectral function in high-$T_c$ superconductors. *Physical Review B* **57**, 4 (1998).

[47] Yi, M. et al. Symmetry-breaking orbital anisotropy observed for detwinned Ba(Fe$_{1-x}$Co$_x$)$_2$As$_2$ above the spin density wave transition. *Proceedings of the National Academy of Sciences of the United States of America* **108**, 6878-6883 (2011).

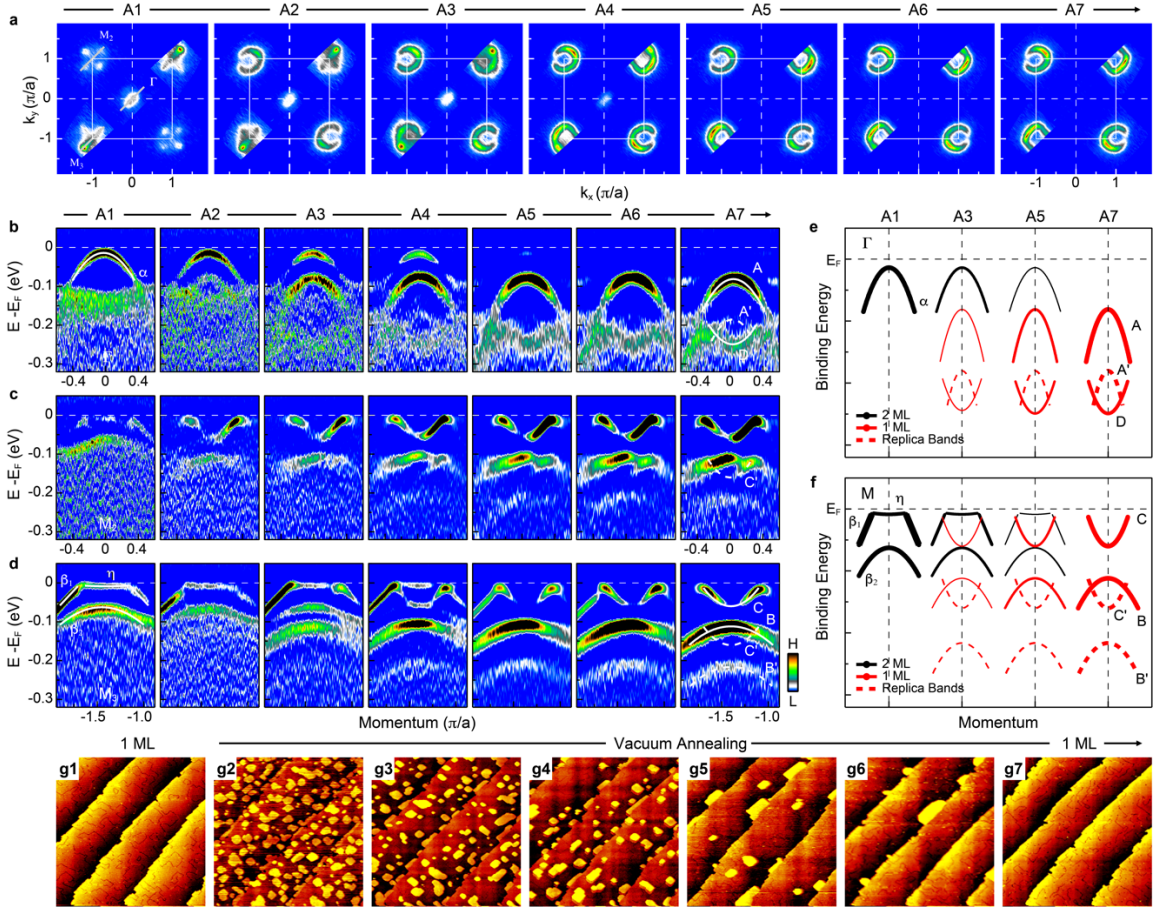

**FIG. 1: Fermi surface and band structure evolution of the single-layer FeSe/STO film during vacuum annealing.** This sample contains ~20% excess Fe [35]. **(a)** Fermi surface at different annealing sequences. The images are obtained by integrating the spectral weight over an energy window [-0.02, 0.01] eV with respect to the Fermi level. For convenience, we use A1-A7 to denote annealing sequences. The annealing condition for each sequence is illustrated in Supplementary Fig. S1a. **(b-d)** Band structure at different annealing sequences for the momentum cuts across Γ (0,0) **(b)**, $M_2$ (-π,π) **(c)** and $M_3$ (-π,-π) **(d)**. The images are second derivative of the original data with respect to the energy, and the original bands are shown in Supplementary Fig. S2. The location of the three cuts are marked in the left-most panel of **(a)**. **(e)** Schematic illustration on the evolution of the two sets of bands around Γ point for A1, A3, A5 and A7 sequences. The black lines represent one set of bands while the red lines represent the other. The line width of bands reflects their intensities. **(f)** Same as **(e)** but for band structures around M point. **(g1-g7)** show STM topographies (300nm×300nm) of a single-layer FeSe/STO film. **(g1)** STM image of the initial perfect single-layer FeSe/STO film prepared at 490°C and annealed at 530°C for 4 hours. Then the sample was deposited Se on the surface at room temperature and annealed in six consecutive sequences at different temperatures and times. The condition for each annealing sequence is illustrated in Supplementary Fig. S1b. STM images **(g2-g7)** were taken after each annealing sequence. Yellow patches in **(g2-g6)** represent the second FeSe-layer formed by the reaction of Se with the excess Fe [35]. The second-layer FeSe islands grow in size and their total amount decreases with annealing **(g2-g6)**. It completely disappears after sequence 7 and the sample recovers to a perfect single-layer FeSe/STO film **(g7)**.

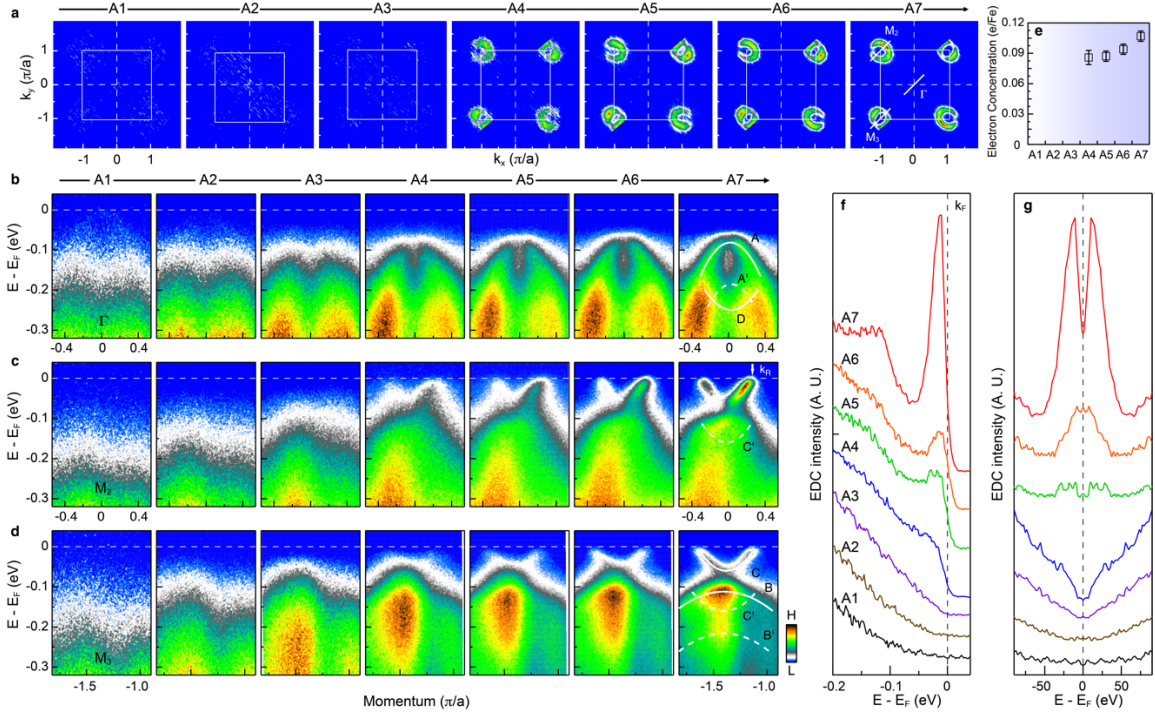

**FIG. 2: Fermi surface and band structure evolution of the pure single-layer FeSe/STO film during annealing.** The sample contains negligible amount of excess Fe (see Fig. S4 in Supplementary). **(a)** Fermi surface for the pure single-layer FeSe/STO film at different annealing sequences A1-A7. The annealing condition for each annealing sequence is shown in Supplementary Fig. S1c. The images are obtained by integrating the spectral weight over an energy window [-0.02, 0.01] *eV* with respect to the Fermi level. **(b-d)** Band structure at different annealing sequences for the momentum cuts across Γ **(b)**, $M_2$ **(c)** and $M_3$ **(d)**. The location of the three cuts are marked in the right-most panel of **(a)**. We note that the weak signal of a hole-like band in the initial sample (left-most panel of **(b)**) is from the tiny remnant second-layer FeSe islands; it disappears after the annealing sequence A2. **(e)** Electron concentration for each annealing sequence estimated from the area of the electron pockets near M point. **(f)** EDCs at the Fermi momentum $k_R$ (marked by arrow in right-most panel in **(c)**) for different annealing sequences. The corresponding symmetrized EDCs are shown in **(g)**; EDC symmetrization is a standard procedure to detect a gap opening [46]. In **(f)**, with increasing electron doping, the spectral weight near the Fermi level gets stronger and finally forms a well-defined peak (A7). In **(g)**, the gap observed for sequences A1-A5 is an insulating gap which does not change its size with temperature while the gap for the sequence A7 is a superconducting gap [27].

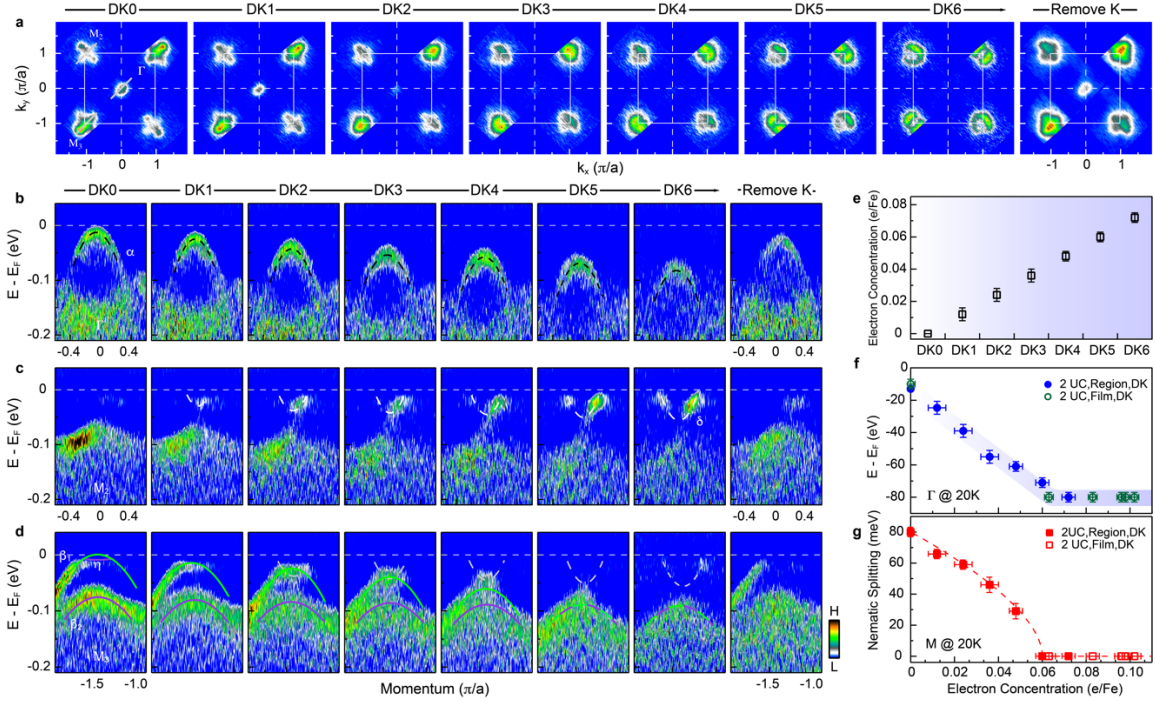

FIG. 3: **Fermi surface and band structure evolution of the double-layer FeSe regions on the single-layer FeSe/STO film with sequential potassium depositions.** The initial sample is similar to the one in Fig. 1g2; it contains ~20% second FeSe-layer due to Se deposition and its reaction with the excess Fe during annealing. As explained in the main text, the signals observed here are mainly from the double-layer FeSe regions because the single-layer FeSe is insulating in this doping range and contributes little signal in the measured energy range. **(a)** Fermi surface at different deposition sequences. The images are obtained by integrating the spectral weight over an energy window [-0.02, 0.01] *eV* with respect to the Fermi level. We use DK0-DK6 to denote the deposition sequences, where DK0 represents the initial sequence without deposition while for each sequence of DK1-DK6, the deposition is 1 minute with keeping the sample at 20 *K*. After the potassium deposition DK6, the sample was warmed up and stayed at 200°C for 2 hours to remove potassium. **(b-d)** Band structure at different potassium deposition sequences for the momentum cuts across Γ **(b)**, $M_2$ **(c)** and $M_3$ **(d)**. The location of the three cuts are marked in the right-most panel of **(a)**. The images are the second derivative of the original data with respect to the energy, and the original bands are shown in Supplementary Fig. S5. For convenience, the bands are labeled as α, $β_1$, $β_2$, η and δ in the left-most panels of **(b-d)**. The electronic structure of the sample (right-most panels in a-d) recovers to its initial state after the potassium is removed, indicating that its evolution with potassium deposition and evaporation is fully reversible. **(e)** Electron concentration determined from the electron-like Fermi surface area near M point. **(f)** The energy position variation of the hole-like α band top at Γ point as a function of electron concentration. The data are from **(b)** and double-layer FeSe/STO films (Fig. S9 in Supplementary). **(g)** Band splitting at M point as the function of electron concentration. The nematicity causes $d_{yz}$ band ($β_1$) and $d_{xz}$ band ($β_2$) splitting near M point [39,47]. The splitting size is obtained by the energy position difference between the top of the two bands.

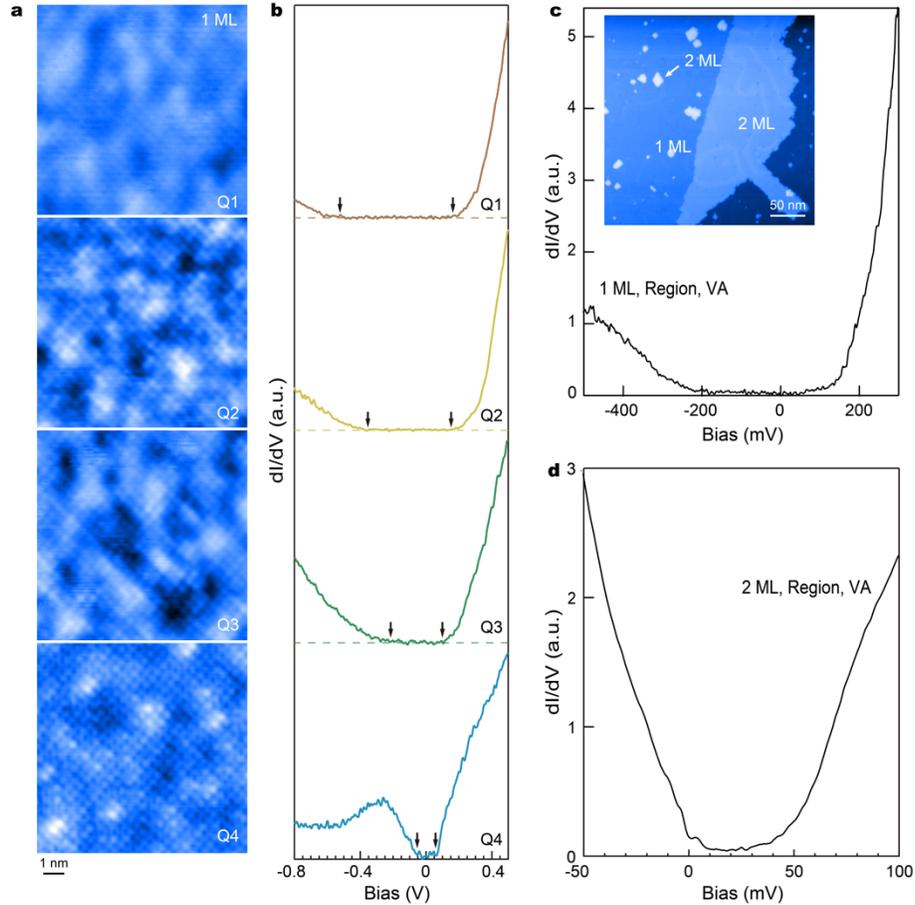

**FIG. 4: The insulating gap evolution with the increase of electron doping in single- layer FeSe/STO film and metallic behavior in double-layer FeSe/STO film revealed by STM/STS.** The sample preparation and the vacuum annealing process mimic those in Fig. 1. The initial perfect single-layer FeSe/STO film was prepared at 450°C and annealed at 530°C. Then the sample was deposited Se on the surface at room temperature and annealed in four consecutive sequences Q1-Q4: 300°C for 2 hours (Q1), 350°C for 2 hours (Q2), 400°C for 2 hours (Q3) and 480°C for 2 hours (Q4). **(a)** Atomically resolved STM images of the single-layer FeSe regions for the four annealing sequences measured at 4.2 *K*. The corresponding STS curves are shown in **(b)**. The arrows in each spectrum indicate the range of the insulating gap. **(c-d)** Detailed dI/dV curves taken at 4.2 *K* on the single-layer region and the double-layer region that coexist on the same sample as Q3 in **(a)** (as shown in the inset of **(c)**). While the spectral weight is zero at zero bias for the single-layer region **(c)**, the double-layer region has considerable amount of spectral weight at the Fermi level.

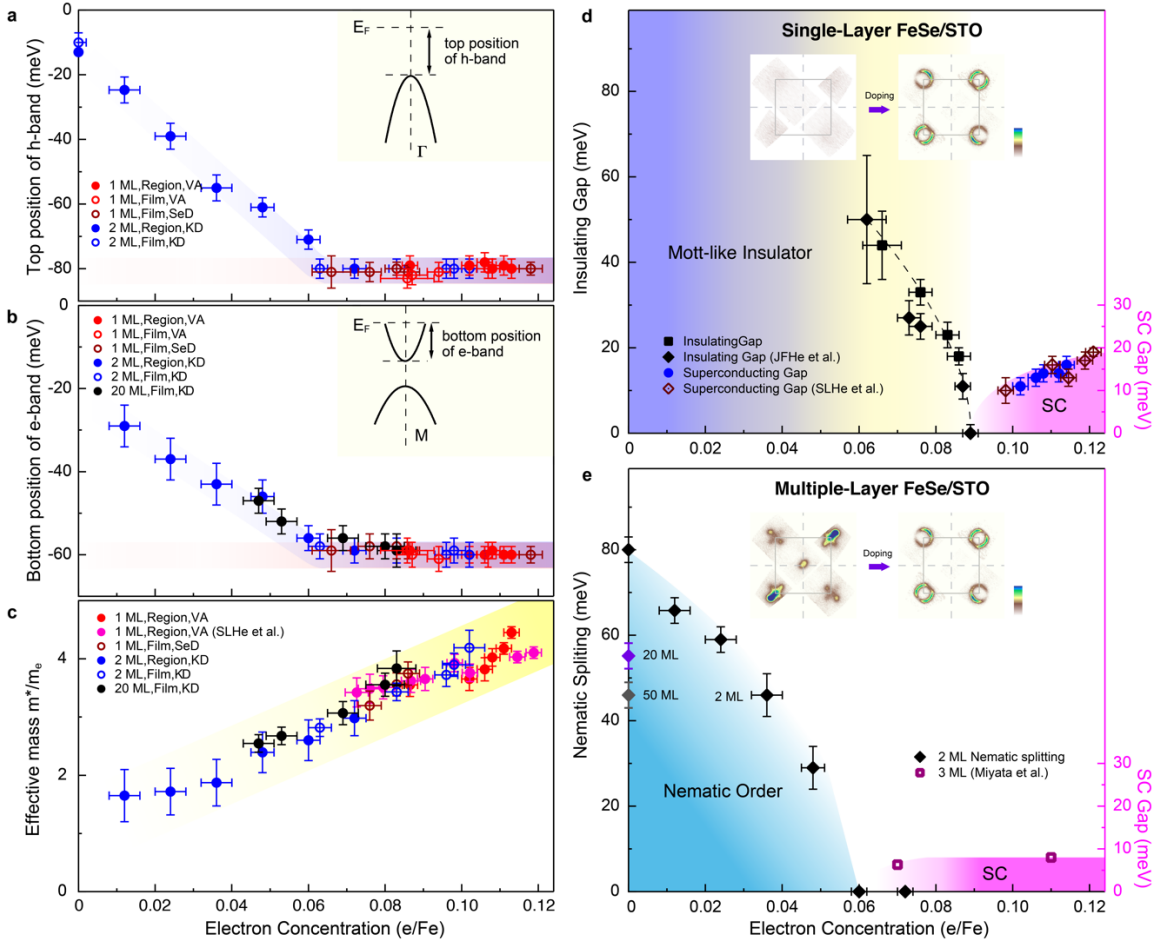

**FIG. 5: Quantitative band evolution with doping and electronic phase diagrams for the single-layer and multiple-layer FeSe/STO films. (a)** Variation of the band top position for the hole-like band near Γ point as a function of electron doping. **(b)** Variation of the band bottom position for the electron-like band near M point as a function of electron doping. The band top and band bottom positions for Γ and M are shown schematically in insets of **(a)** and **(b)**, respectively. **(c)** Effective mass $m^*$ extracted from the electron-like bands crossing M points. $m_e$ refers to the free-electron mass. **(d)** Phase diagram of the single-layer FeSe/STO film that shows a Mott-like insulator phase at low doping side and a superconducting phase with increasing electron concentration. The parent phase is insulating. The insulating gap shrinks with increasing doping and vanishes near ~0.09 e/Fe where there is an insulator-superconductor crossover [27]. Further increase of electron doping leads to an increase of the superconducting gap size and the corresponding $T_c$ [18]. The figure also includes some data from the references [27] and [18]. The top inset shows schematically the Fermi surface topology change from insulating region to superconducting region. **(e)** Phase diagram of the multiple-layer FeSe/STO films. The parent phase is nematic as evidenced by the band splitting of the $d_{yz}$ and $d_{xz}$ bands near M point. The electron doping causes suppression of the nematic order and its disappearance at a critical value that is ~0.06 e/Fe for the double-layer. When the doping level is beyond the critical doping, the Fermi surface only consists of electron pockets near M points. Some data are adopted from the references [42, 43]. The top inset shows schematically the Fermi surface topology change from the parent nematic state to the superconducting state. ML refers to monolayer in the figure.